\documentclass[aps, prb, reprint, superscriptaddress]{revtex4-2}

\usepackage{amsmath}
\usepackage{newtxtext, newtxmath}
\usepackage{braket}
\usepackage{graphicx}
\usepackage{color}
\definecolor{bblue}{rgb}{0, 0.0, 0.8}
\definecolor{rred}{rgb}{0.7, 0.0, 0.0}

\usepackage{hyperref}
\hypersetup{
    colorlinks=true,        % false: boxed links; true: colored links
    linkcolor=rred,          % color of internal links (change box color with linkbordercolor)
    citecolor=rred,        % color of links to bibliography
    filecolor=rred,      % color of file links
    urlcolor=rred           % color of external links
}

\begin{document}
% Title

%updated 20250702 ikeda
\title{Pulse magnet of 10 T for power laser experiments with x-ray free-electron laser diagnostics}

%updated 20250702 ikeda
\author{Akihiko~Ikeda}
\email[]{a-ikeda@uec.ac.jp}
\author{Kosuke~Noda}
\author{Yutaro~Yamanaka}
\author{Yuma~Urabe}
\author{Keiichiro~Kawai}
\affiliation{Department of Engineering Science, University of Electro-Communications, Chofu, Tokyo 182-8585, Japan}
\author{Yasuhiro~H.~Matsuda}
\affiliation{Institute for Solid State Physics, University of Tokyo, Kashiwa, Chiba 277-8581, Japan}
\author{Hirotaka~Nakamura}
\author{Ryusuke~Yamamoto}
\author{Yoshiki~Naito}
\author{Yasuhiro~Kuramitsu}
\author{Kai~Taketoshi}
\author{Naoki~Yamagata}
\author{Norimasa~Ozaki}
\affiliation{Graduate School of Engineering, The University of Osaka, Suita, Osaka 565-0871, Japan}
\author{Tatiana~Pikuz}
\affiliation{Institute for Open and Transdisciplinary Research in Initiatives, The University of Osaka, 2-6 Yamadaoka, Suita, Osaka, 565-0871, Japan}
\affiliation{Institute of Laser Engineering, The University of Osaka, 2-6 Yamadaoka, Suita, 565-0871, Japan. }
\author{Yoichi~Sakawa}
\author{Takayoshi~Sano}
\author{Ryosuke~Kodama}
\affiliation{Institute of Laser Engineering, The University of Osaka, 2-6 Yamadaoka, Suita, 565-0871, Japan. }
\author{Taichi~Morita}
\author{Tomoya~Ogawa}
\affiliation{Faculty of Engineering Sciences, Kyushu University, 6-1 Kasuga-Koen, Kasuga, Fukuoka 816-8580, Japan}
\author{Kohei~Miyanishi}
\email[]{miyanishi@spring8.or.jp}
\affiliation{RIKEN SPring-8 Center, Sayo, Hyogo 679-5148, Japan}
\author{Toshinori~Yabuuchi}
\affiliation{Japan Synchrotron Radiation Research Institute (JASRI), Sayo, Hyogo 679-5198, Japan}
\author{Rigon~Gabriel}
\author{Bakandreas~Stavros}
\author{Koenig~Michel}
\author{Bruno~Albertazzi}
\email[]{bruno.albertazzi@polytechnique.edu}
\affiliation{LULI, CNRS, \'{E}cole Polytechnique, CEA, Sorbonne Universités, Institut Polytechnique de Paris, F-91128 Palaiseau cedex, France}
\date{\today}

\begin{abstract}
The importance of investigating magnetized plasmas/solids in extreme conditions has grown over the last decades, particularly in the field of high energy density physics (HEDP), such as laboratory astrophysics and inertial confinement fusion. However, up to now, the unique capabilities of an X-ray free-electron laser (XFEL), such as high brilliance and low divergence have never been exploited for this type of research. In this paper, we present the first platform  developed at SACLA, Japan, that combines a high-power optical laser for generating matter under extreme conditions of pressure and temperature, an XFEL probe, and an external magnetic field. The high current is produced using a 2 kV, 4.8 kJ pulsed power system giving a maximum current of 10 kA which is synchronized with the optical laser and XFEL in a vacuum environment. It flows through a split-pair coil to generate a high magnetic field (10 T at 6 kA) which has 1 cm access every 45$^{\circ}$ in the equatorial plane and 90$^{\circ}$ in the poloidal one. This platform offers new opportunities to study high-energy-density matter in strong magnetic fields, including shock propagation, instability growth, and turbulent plasma dynamics.

\end{abstract}

\maketitle

\section{Introduction}
%

%The coupling between high magnetic field and high power lasers has become a significant trend in relevance to laboratory astrophysics \cite{AlbertazziRSI2013, AlbertazziScience2014, Mabey2020} and inertial confined fusion technology \cite{Moody2021}, where one investigates the influence of magnetic fields to the propagation of shock waves in materials and the evolution of laser-produced plasma in magnetohydrodynamic (MHD) turbulence.
The advent in coupling high power laser with external magnetic field has helped to advance in the interpretation of various physical mechanism at play in astrophysics \cite{AlbertazziRSI2013, AlbertazziScience2014, Mabey2020}, in inertial confinement fusion \cite{Moody2021}, or even in planetology where magnetic field could influence phase transitions or opacity, and in condensed matter.
However, usually in such systems, the plasma to be diagnosed is at high density and does not allow us to use conventional optical diagnostics. One option is to generate a laser x-ray source to probe it. These sources are, however, restricted as the maximum spatial resolution is of the order of ~25 $\mu$m at best, that is often too large to visualize all required details.

Over the last few years, a new method for diagnosing dense plasmas with sub-micron spatial resolution using a LiF crystal has been developed at the x-ray free-electron (XFEL) laser facility of SACLA in Japan \cite{IshikawaNP2012} in combination with a nanosecond high power laser \cite{Inubushi2020AS}.
It has recently proved its tremendous capability to resolve spatially systems with less than one-micron resolution associated with a large field of view (FOV) \cite{Faenov2018, Rigon2021}. 
Understanding such transient, non-linear and non-equilibrium phenomena benefits from a high-resolution and a femtosecond high-speed observation using single-shot radiography \cite{Katagiri2023Science}, and diffraction measurements  \cite{AlbertazziSA2017, Katagiri2020PRL}, which are realized by an x-ray free-electron laser (XFEL).
The unique characteristics of the XFEL beam, however, have not yet been fully exploited to investigate magnetized processes in HEDP.
%We want to stress here that it will be the first platform around the world enabling to couple a strong external magnetic field (up to 20 T), high power lasers, and an XFEL beam.
%These developments are already planned at the Eu-XFEL and are currently under discussion at LCLS.
%This is also the only way to strongly magnetize compressed matter or plasmas produced using a high-power laser and utilize an XFEL beam (as a pump or probe).

\begin{figure}
\begin{center}
\includegraphics[width = \columnwidth]{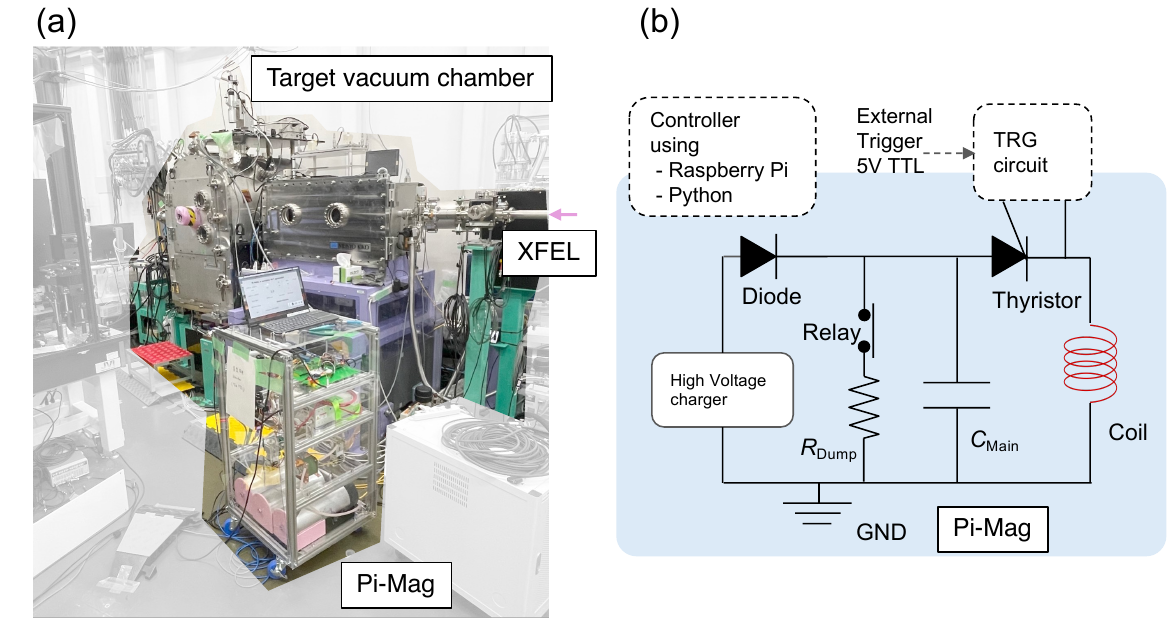}
\caption{
(a) Photography of the platform showing the pulse power system (Pi-Mag), the XFEL beam and the experimental chamber in EH5 at SACLA
(b) Electrical circuit of the Pi-Mag system generating the high current
\label{bank}}
\end{center}
\end{figure}

\begin{figure}
\begin{center}
\includegraphics[width = \columnwidth]{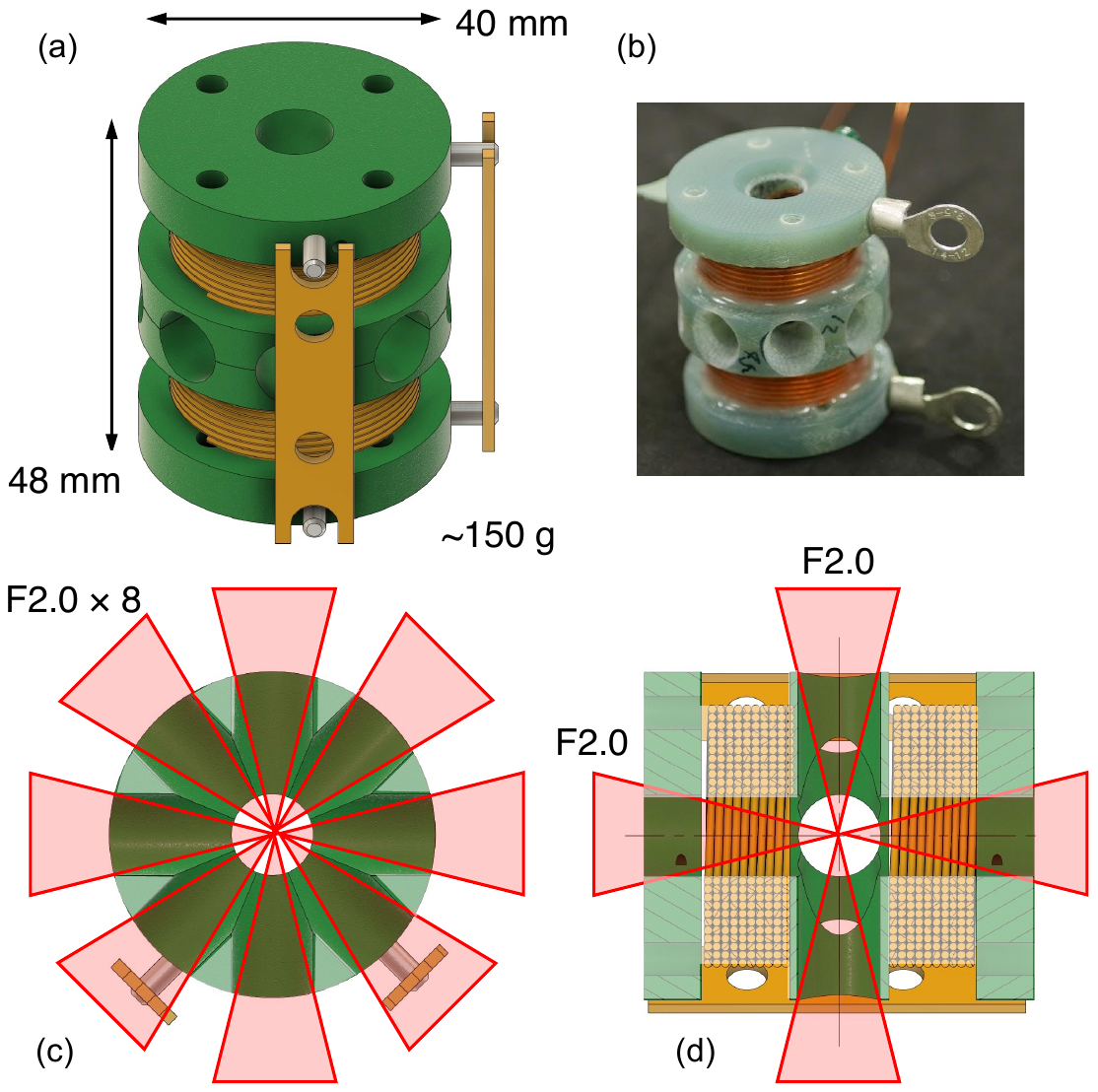}
\caption{
(a) Schematic of the pulse magnet with multiple optical access.
(b) Photo of the pulse magnet with multiple optical access.
(c) A cross-sectional view of the pulse magnet showing transverse optical accesses.
(d) A cross-sectional view of the pulse magnet showing longitudinal optical accesses.
\label{coil}}
\end{center}
\end{figure}

Indeed, ultra-high spatial resolution X-ray imaging of dense magnetized plasmas could unlocks transformative insights across multiple scientific frontiers. By combining sub-micron to micron resolution with a large field of view (FOV), this capability enables groundbreaking investigations, for example into:
(1) the discovery and detailed characterization of novel physical mechanisms driving magnetohydrodynamic (MHD) instabilities,
(2) the influence of magnetic fields on wave dynamics and propagation in solid materials,
(3) The modification of equations of state under the presence of strong magnetic fields,
(4) The behavior of turbulent plasmas in magnetized environments, including the study of turbulent MHD phenomena.

%The present platform extends previous XFEL–LiF radiography experiments at SACLA from unmagnetized laser-driven hydrodynamics to externally magnetized high-energy-density systems.
%Obtaining Ultra-High Spatial X-ray Imaging of a dense magnetized plasma is beneficial for a huge number of scientific interests, where the sub-micron to micron spatial resolution coupled to a large FOV will reveal:
%(1) New physical mechanisms occurring in MHD instabilities and their detailed studies,
%(2) How the dynamics and propagation of waves in solids are affected by a magnetic field,
%(3) How the equation of state is affected by a magnetic field,
%(4) How a turbulent plasma is affected by a magnetic field (turbulent MHD).
In addition, this platform can be coupled to various x-ray diagnostics capabities such as a Talbot-Lau interferometer to measure changes in the plasma's complex refractive index and ionization state induced by the magnetic field. This is useful for understanding how pressure and/or magnetic field modifies the electron charge distribution and the intra-atomic/molecular distance.

Here, we present an experimental platform developed at the EH5 of SACLA enabling us to couple the high power laser, the XFEL beam and a magnetic field of 10 T, homogenous over a volume of 0.5 cm$^{3}$, corresponding to a magnetic pressure of $\sim0.04$ GPa. The repetition rate of the magnetic-field pulse is 10 min/pulse and authorized us to shoot 10 targets before venting the experimental chamber.
This paper is structured as follows. Section II introduces the magnetic field platform, detailing the pulse power system (Pi-Mag), the split-pair coil design, the experimental chamber assembly, and the synchronization of discharges with the XFEL and high-power optical beams. Section III presents preliminary experimental results, including an example of MHD turbulence diagnostics where a magnetic-field-induced change in the power spectrum slope was observed.

%The paper is organized as follows: the first part describes the magnetic field platform such as the pulse power system (Pi-Mag), the coil, the general assembly inside the experimental chamber and typical discharges with the XFEL and high power optical beams. The second part illustrates an example of data obtained in the context of MHD turbulence, where we could observe a change in the slope of the power spectrum due to magnetic effects. 

%We obtained a preliminary observation of the power-laser-induced shockwave in a soft form using XFEL radiography, with and without a magnetic field.

\begin{figure}
\begin{center}
\includegraphics[width = \columnwidth]{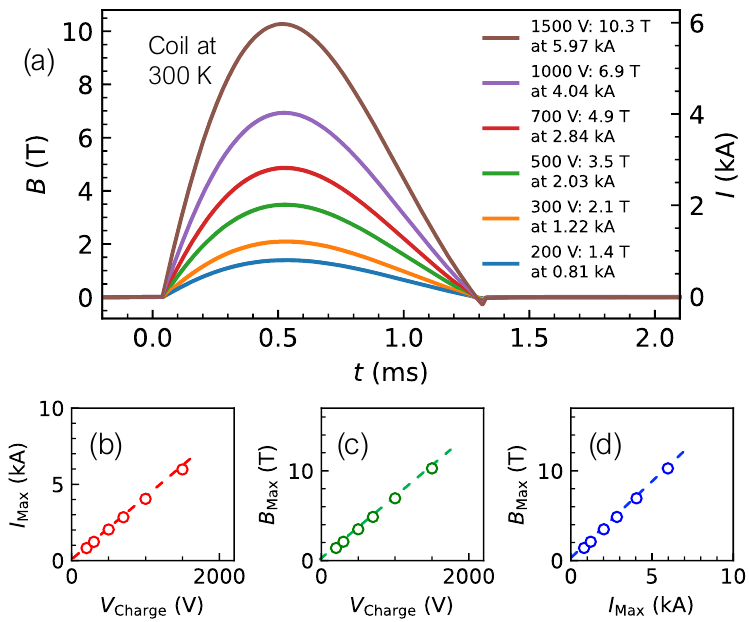}
\caption{
(a) Magnetic field profile of the pulse magnet at room temperature.
(b) The relationship between the charging voltage and the maximum current.
(c) The relationship between the charging voltage and the maximum magnetic field.
(d) The relationship between the generated current and the generated magnetic field.
\label{pulse}}
\end{center}
\end{figure}

\begin{figure}
\begin{center}
\includegraphics[width = 0.7\columnwidth]{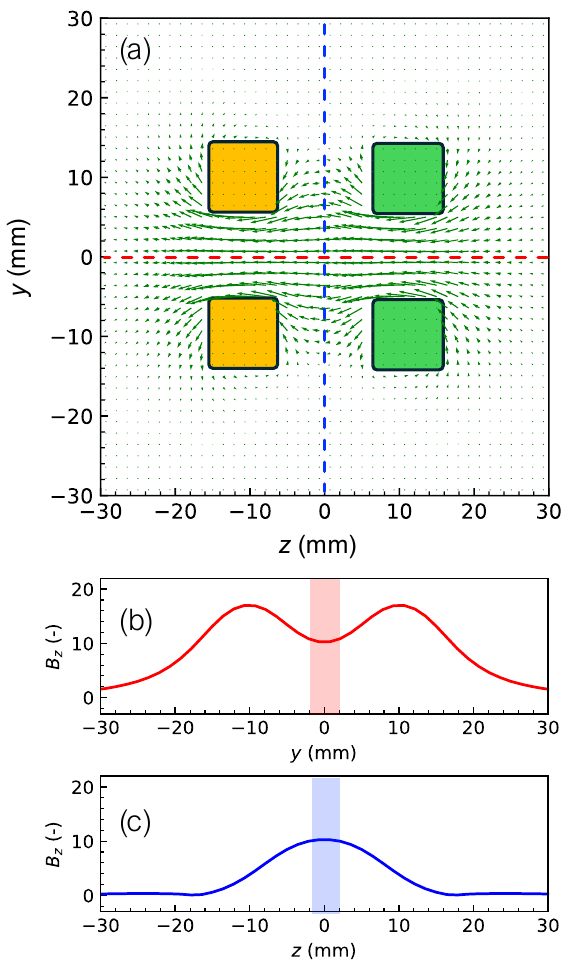}
\caption{
(a) The spatial distribution of the magnetic field on the $y-z$ plane, where $y$ axis is the coil axis.
(b) The profile of $B_z$ along the red dashed line in (a).
(c) The profile of $B_z$ along the blue dashed line in (a).
\label{dist}}
\end{center}
\end{figure}

\begin{figure*}
\begin{center}
% \includegraphics[width = 0.8\textwidth]{./figs/design2.pdf}
% \caption{
% (a) The schematic view of the experimental configuration of the coil, power laser, and XFEL.
% (b) The schematic view of the experimental configuration of the coil in the vacuum chamber with the power laser, the XFEL, LiF, and the scintillator-based CCD \cite{KameshimaOL2019}.
% (c) The schematic view of the experimental configuration of the cross-section of the coil with the magnetic field direction and the power laser.
% \label{design2}}
\includegraphics[width = 0.8\textwidth]{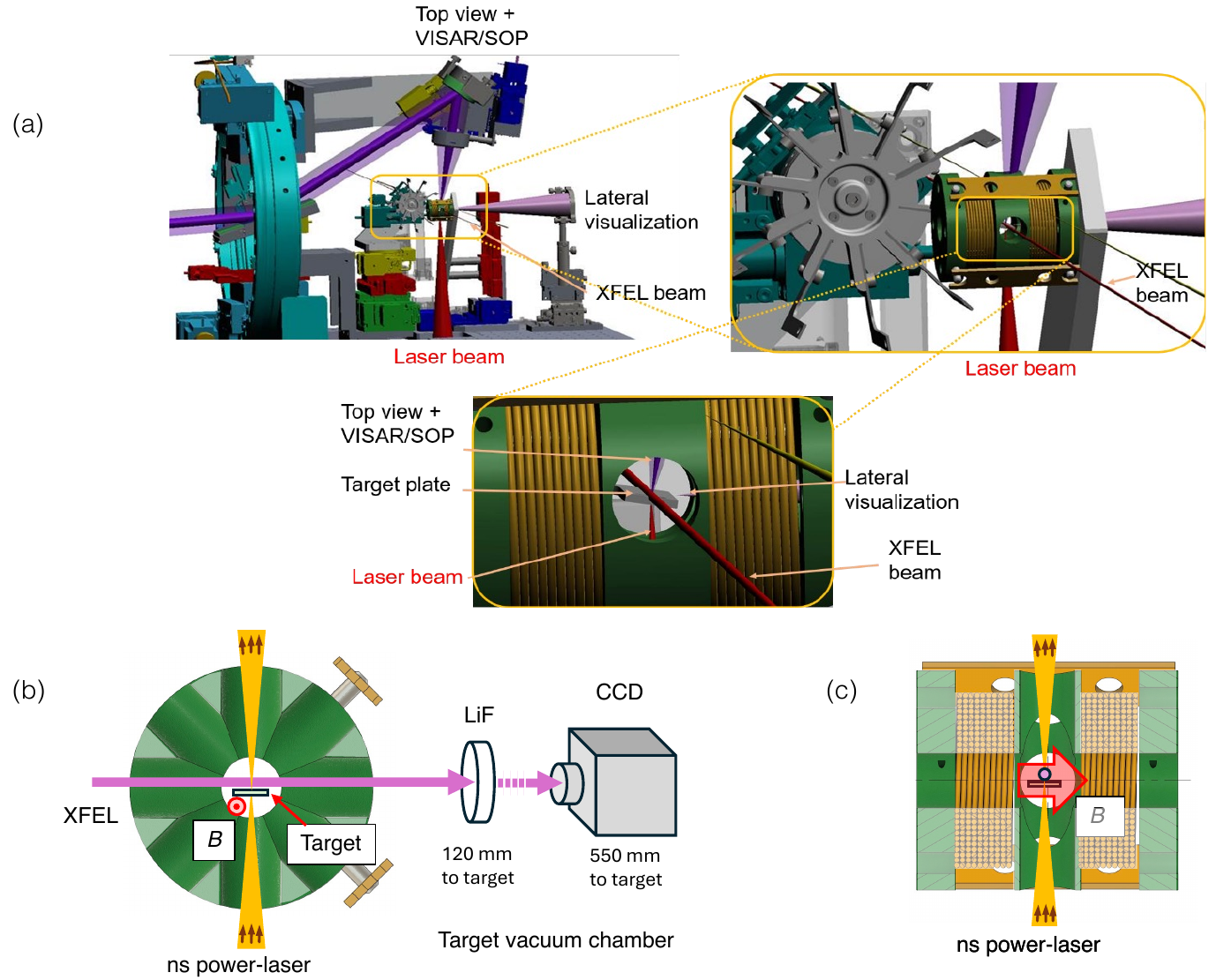}
% Akihiko May 13
% How about using your figure including the sapmle feeding mechanics?
% I imported it from your proposals.
% It has VISOR/SOP that is not used in the last experiment but I think it is OK.
\caption{
(a) The visualized layout of the experimental configuration.
The sample feeding mechanics, the coil, power laser, and XFEL in the experimental target chamber in EH5 of BL3 in SACLA are shown.
(b) The schematic view of the experimental configuration of the coil in the vacuum chamber with the power laser, the XFEL, LiF, and the scintillator-based CCD \cite{KameshimaOL2019}.
(c) The schematic view of the experimental configuration of the cross-section of the coil with the magnetic field direction and the power laser.
\label{design2}}
\end{center}
\end{figure*}

\begin{figure}
\begin{center}
\includegraphics[width = \columnwidth]{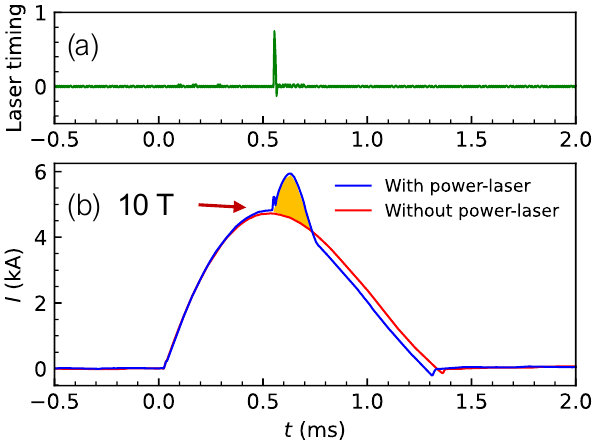}
\caption{
(a) The timing of the power laser pulse
(b) The profile of the magnetic field pulse with and without the power laser.
\label{shot}}
\end{center}
\end{figure}

\begin{figure*}
\begin{center}
\includegraphics[width = 0.8\textwidth]{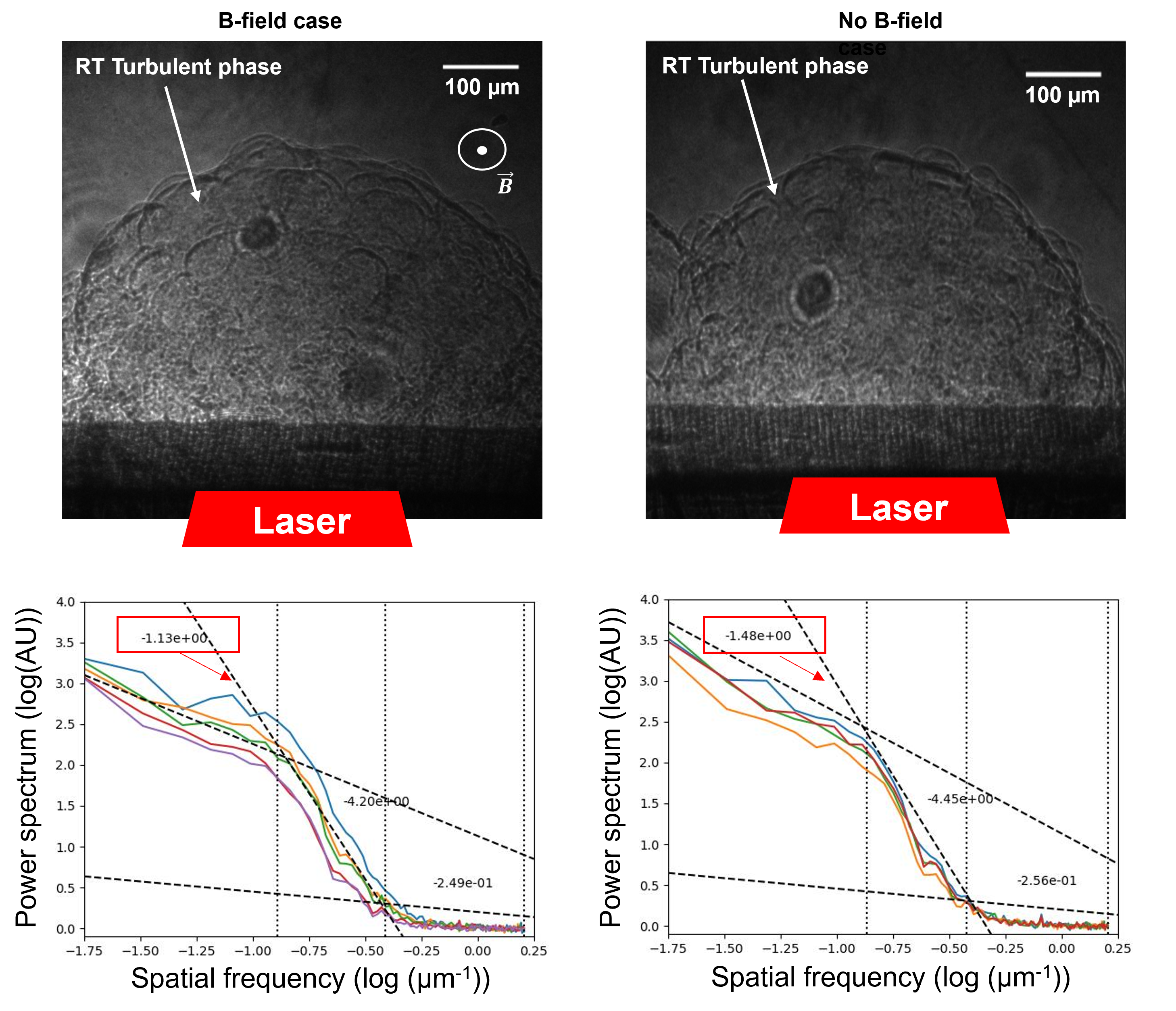}
\caption{
Spatial power spectrum, 
left: with a B-field perpendicular to the propagation of the shock:
right, without B-field. 
The top images are the x-ray radiography of a target exposed in \cite{Rigon2021}. The methodology of analysis is explained in details in \cite{Rigon2021} and the associated power spectrum are displayed below each X-ray radiography. The square red rectangle show the difference observed in the power spectrum exponent. 
\label{result}}
\end{center}
\end{figure*}

\section{Instrumentation}
This section describes the different element of the magnetic platform as well as typical discharge used inside EH5. 

\paragraph{Pulse power of the pulse magnet}
%The pulse power (Pi-Mag) is composed of a capacitor 2.4 mF and an HV charger at 2000 V with a stored energy of 4.8 kJ Figs. \ref{bank}(a) and \ref{bank}(b).
%Pi-Mag weighs only 50 kg, making implementation and removal very easy per beam time.
%By triggering the thyristor, it is possible to obtain a maximum discharge current of 10 kA injected into the split-pair coil presented in Figs. \ref{coil}(a)-\ref{coil}(d).
%The high current is produced using a pulse power system outside of the target chamber but inside the experimental hatch for the hard X-ray equipped with an interlock system synchronized with the door of the experimental hatch, ensuring the safety of humans.
%Pi-mag, which is controlled from outside the experimental hatch using a VNC connection to the Raspberry Pi that is the controller for the Pi-Mag \cite{IkedaJAP2024}.
The Pi-Mag pulse power system consists of a 2.4 mF capacitor and a 2000 V high-voltage charger, storing a total energy of 4.8 kJ (Figs. \ref{bank}(a) and \ref{bank}(b)). With a compact design weighing only 50 kg, Pi-Mag can be easily installed and removed between beam sessions.
By triggering the thyristor, the system delivers a maximum discharge current of 10 kA to the split-pair coil (Figs.  \ref{coil}(a)-\ref{coil}(d)). The pulse power system is housed outside the target chamber but inside the experimental hutch for hard X-ray experiments. It features an interlock system synchronized with the hutch door, ensuring operational safety. Pi-Mag is remotely controlled from outside the experimental hatch via a VNC connection to a Raspberry Pi controller \cite{IkedaJAP2024}.

\paragraph{Split coil}
%The coil is a split-pair coil with a 10 mm bore.
%A wire of CuAg alloy (Ag 10 \%) of $\phi1.0$ mm with an AIW coating.
%Each coil is wound with 10 turns and 10 layers, with a split of 12 mm.
%To fix the wire, we used the epoxy Stycast 1266, mixing the glass powder at a 1:1 weight ratio.
%The two coils are electrically connected in parallel.
%There are optical accesses of 8 mm diameter in the equatorial plane every 45$^{\circ}$ and two accesses at 90$^{\circ}$ in the polar plane. (see Figs. \ref{coil}(c) and \ref{coil}(d)).
%The total length of the coil is of $\phi\sim48$ mm in the magnetic field axis and of $\phi\sim40$ mm diameter perpendicular to the magnetic field without the connectors.
%It weighs only 150 g, making it easy to hold and position in the vacuum chamber.
The split-pair coil features a 10 mm bore and is constructed from a CuAg alloy wire (10 \% Ag) with a $\phi1.0$ diameter and an AIW coating. Each coil consists of 10 turns and 10 layers, with a 12 mm split between the pair. The wire is secured using Stycast 1266 epoxy, mixed with glass powder in a 1:1 weight ratio. The two coils are electrically connected in parallel.
Optical access is provided via 8 mm diameter ports spaced every 45 $^{\circ}$ in the equatorial plane and two ports at 90$^{\circ}$ in the polar plane (Figs. \ref{coil}(c) and \ref{coil}(d))). The coil’s total dimensions are $\phi\sim48$ mm along the magnetic field axis and $\phi\sim40$ mm perpendicular to it (excluding connectors). With a total weight of 150 g, the coil is lightweight and can be easily positioned within the vacuum chamber.

\paragraph{Magnetic field profile}
As shown in Fig. \ref{pulse}, the coil produces 10 T with 6 kA and a pulse duration of 1.3 ms.
The heating of the coil limits the repetition to 10 T every 10 min.
The homogeneity of the magnetic field at the coil center is $\pm 2$ mm, which is sufficiently larger than the FOV ($<0.6$ mm) of the diagnostics (See Fig. \ref{dist}).

\paragraph{The magnetic field pulse with power laser}
As shown in Figs. \ref{shot}(a) and \ref{shot}(b), the XFEL and the high-power laser pulse are synchronized with the maximum of the magnetic field pulse.
We noticed that our coil never escapes from an abnormal discharge in the vacuum chamber, only when the power laser is accompanied.
The abnormal discharge occurs because the electrical insulation between the two electrodes of the coil becomes significantly reduced due to the generated plasma and outgassing from the power laser striking the target.
We successfully suppressed the abnormal discharge at the coil in the target vacuum chamber by applying a large amount of electrical insulation tape to every electrode in the target vacuum chamber.

\paragraph{High Power laser and XFEL pulse}
The High-power nanosecond laser ($\Delta t = 5$ ns)  at $\lambda = 532$ nm with a pulse energy of 15-20 J is used, where the beam size is 250 $\mu$m, and the power density is $1.5\times10^{13}$ W cm$^{-3}$.
The XFEL pulse is a non-focused beam of $\phi 0.6$ mm at $E = 7$ keV with $\Delta E/E \sim 1/1000$.
The XFEL pulse is delayed by a few to a few hundred nanoseconds relative to the power laser.

% \paragraph{Main diagnostics}
% It is composed of a LiF crystal and we use the XFEL beam as a probe (see Fig. \ref{design2}) \cite{Faenov2018, Rigon2021}.
% The intrinsic resolution of this system depends on the spacing of the colour centres within the LiF (a few nanometres), the resolution of the confocal microscope used to read out the data (0.26 $\mu$m), and the secondary electron avalanche and diffusion of colour centres.
% In this experiment, the image contrast is ensured initially by the difference of density of the different materials involved (initially the foam is more than ten times lighter than the pusher), then by the difference in the absorption coefficients of those materials (the high Z material, Br, absorbs more x-rays), and by phase-contrast (PC) effects.
% This diagnostics will enable us to determine the power spectrum of the turbulent flow over almost 2 orders of magnitude (from the injection length corresponding more or less to the focal spot of the optical laser to the $\mu$m level using LiF).

\paragraph{XFEL radiography diagnostics}
The diagnostic system consists of a LiF crystal detector and an online X-ray imaging detector, using the XFEL beam as a probe, as shown in Fig. 5. The LiF crystal is the primary detector for high-resolution radiographic imaging and for the subsequent analysis of spatial structures and turbulent power spectra. At a photon energy of 7 keV, this LiF-based diagnostic has been demonstrated to provide a spatial resolution of approximately 0.6 $\mu$m in SACLA XFEL radiography experiments \cite{Rigon2021}. After XFEL irradiation, the image recorded in the LiF crystal as colour-centre distributions is read out offline using a confocal microscope. The effective spatial resolution is determined by several factors, including the colour-centre distribution in LiF, the resolution of the confocal microscope, and broadening associated with secondary-electron transport and colour-centre diffusion.

In parallel, the CCD-based X-ray imaging detector is used as an online diagnostic. It provides immediate feedback during the experiment, allowing visualization of the radiographic image, confirmation of the XFEL–target alignment, monitoring of shot conditions, and control of the experimental setup. However, the CCD detector is not used for the highest-resolution image analysis or for the quantitative extraction of turbulent spectra.

In the present experiment, radiographic contrast is produced by a combination of density differences between the target materials, absorption contrast, and phase-contrast effects. %Initially, the foam has a density more than an order of magnitude lower than that of the pusher. In addition, the brominated plastic pusher contains a high-Z element, Br, which increases X-ray absorption relative to the surrounding material. The LiF-based high-resolution radiographs therefore enable analysis of the spatial structure of the unstable flow and extraction of its power spectrum over nearly two orders of magnitude, from the injection scale, comparable to the optical-laser focal spot size, down to micron-scale structures.

%paragraph{Target}
%This is a multi-layer target composed of a 10 $\mu$m parylene ablator, a 40 $\mu$m modulated brominated plastic (C$_{8}$H$_{7}$Br) pusher and a resorcinol formaldehyde (C$_{15}$H$_{12}$O$_{4}$) foam (100 mg cm$^{3}$).
% The optical laser of SACLA deposits $\sim15-20$ J on a $\sim240$ $\mu$m diameter focal spot, which is smoothed by a hybrid phase plate (HPP), in a $\sim5$ ns square pulse giving an intensity in the range $\sim0.5 - 1 \times 10^{13}$ W cm$^{2}$ on target.
%This leads to the ablation of the first layer, the ablator, and in reaction generates a shock wave inside the solid target.
%The shock wave, during its propagation, puts into motion the rest of the target, in particular the pusher that expands into the foam and decelerates.
%This situation is Rayleigh–Taylor unstable.
%Most of the pushers will be modulated in order to make this experiment reproducible and easy to simulate.
%Indeed, this allows the control over which spatial modes will be favored during Rayleigh–Taylor development.
%These modes will not be created randomly due to thermal noise, material roughness, imperfection in the spatial deposition of the laser energy, or numerical noise, in the case of simulation.
%Two kinds of modulation will be used: a simple sine wave (mono-mode), with 40 $\mu$m wavelength and 5 $\mu$m amplitude; the sum of two sine waves (bi-mode), the previous one and another with 15 $\mu$m wavelength and 5 $\mu$m amplitude. 
 
 \paragraph{Outlook for 20 T}
 The magnetic field strength should be increased to at least 15 T by reducing the optical access, which is possible without changing the optical setup in the present experiment.
 Above 20 T, we need to cool the coil with liquid nitrogen through the thermal conduction of the CuAg wire.
 % May be we can aim for 20 T (or at least 15 T) by making a less optically accessible coil like .... I want to make such a coil and just show the test result...
 
\section{Preliminary results}
Producing a complete description of MHD turbulence has previously been described as one of the major unsolved problems in space plasma physics \cite{Goldstein2001}. As such, this topic is one of the most active research areas in astrophysics and space physics today, with large numbers of observational \cite{Tu1995}, theoretical \cite{Sridhar1994} \cite{Goldreich1995} and computational \cite{Vishniac2000} studies published daily. It is, however, perhaps unsurprising that this subject remains mysterious despite the great amount of effort dedicated to its study, as it represents a combination of two phenomena, which themselves are not fully understood within this context: magnetic fields and turbulence. The platform described here opens up a new field of investigation in this domain, as it enables the experimental characterization of magnetized turbulence. 

Indeed, recently \cite{Rigon2021} showed that it is possible to measure the power spectrum of a turbulent flow generated through the development of late phase Rayleigh-Taylor (RT) instabilities. To achieve this goal a multi-layer target composed of a 10 $\mu$m parylene ablator, a 40 $\mu$m modulated (a simple sine wave (mono-mode), with 40 $\mu$m wavelength and 5 $\mu$m amplitude) brominated plastic (C$_{8}$H$_{7}$Br) pusher and a resorcinol formaldehyde (C$_{15}$H$_{12}$O$_{4}$) foam (100 mg cm$^{3}$) is used. The optical laser of SACLA deposits $\sim15-20$ J on a $\sim240$ $\mu$m diameter focal spot, which is smoothed by a hybrid phase plate (HPP), in a $\sim5$ ns square pulse giving an intensity in the range $\sim0.5 - 1 \times 10^{13}$ W cm$^{2}$ on the target. This leads to the ablation of the first layer, the ablator, and in reaction generates a shock wave inside the solid target. The shock wave passes through the modulated layer and initiate the development of RT instability. At later stages, secondary instabilities such as Kelvin-Helmholtz emerge at the tips of the RT fingers, mixing the pusher material with the foam. Upon saturation, a turbulent flow develops, which can be probed across two orders of magnitude using the XFEL beam coupled with a LiF crystal. The conclusion of this study is that it is possible to determine accurately (up to the µm scale) the power spectrum of a non-magnetized turbulent flow testing the validity of the Kolmogorov model.

Here we will extend this study to magnetized turbulent flow to discriminate the different model used in such conditions (weak and strong MHD turbulence) using the same target assembly as described above \cite{Rigon2021}. Fig. 7 presents two differents shots, one without B-field and the second one with a 9.3 T B-field perpendicular to the propagation of the optical laser.   

We could clearly observed a change in the slope of the power spectrum from -1.5 without B-field to -1.1 in the case of a B-field (see red boxes in Fig. 7). This tends to show that the B-field effectively acts on the dissipation of energy and influence strongly the evolution of the system. It suggests that less energy is being transferred to smaller scales.It could be due to the fact that Lorentz force, which opposes the motion of charged particles in the flow inhibits vortex stretching and eddy formation, reducing energy transfer across scales and/or dissipates kinetic energy into magnetic energy or heat (via Ohmic dissipation or magnetic reconnection).

Before being concluding, it is important to characterize the influence of the B-field parrallel to the optical laser axis as B-field can, as well, introduce anisotropic turbulence. This will help us to discriminate the different theoretical models at play (see for example \cite{Schekochihin2022} for a review).  

\section{Summary}
We established a platform for a high-power laser experiment with 10 T diagnosed by an XFEL pulse.
We performed a preliminary experiment at EH5 in SACLA.
The influence of the magnetic field on the propagation of the shockwave is observed, which may originate from magnetic-field effects during dissipation.

\begin{acknowledgements}
The experiment was carried out in SACLA with the approval of JASRI (Proposal No. 2025A8045).
The authors would like to acknowledge the support from the technical staff of the SACLA facility.
They also appreciate the support from the SACLA/SPring-8 Basic Development Program (Albertazzi and Ikeda 2023 - 2024).
This work is partially supported by JST FOREST program No. JPMJFR222W, JSPS Grant-in-Aid for Scientific Research on Innovative Areas (A) (1000 T Science) 23H04861, 23H04859, (C) 24K06988, and MEXT LEADER program No. JPMXS0320210021. 
\end{acknowledgements}

\bibliography{sacla_laser_b}

\begin{thebibliography}{19}%
\makeatletter
\providecommand \@ifxundefined [1]{%
 \@ifx{#1\undefined}
}%
\providecommand \@ifnum [1]{%
 \ifnum #1\expandafter \@firstoftwo
 \else \expandafter \@secondoftwo
 \fi
}%
\providecommand \@ifx [1]{%
 \ifx #1\expandafter \@firstoftwo
 \else \expandafter \@secondoftwo
 \fi
}%
\providecommand \natexlab [1]{#1}%
\providecommand \enquote  [1]{``#1''}%
\providecommand \bibnamefont  [1]{#1}%
\providecommand \bibfnamefont [1]{#1}%
\providecommand \citenamefont [1]{#1}%
\providecommand \href@noop [0]{\@secondoftwo}%
\providecommand \href [0]{\begingroup \@sanitize@url \@href}%
\providecommand \@href[1]{\@@startlink{#1}\@@href}%
\providecommand \@@href[1]{\endgroup#1\@@endlink}%
\providecommand \@sanitize@url [0]{\catcode `\\12\catcode `\$12\catcode
  `\&12\catcode `\#12\catcode `\^12\catcode `\_12\catcode `\%12\relax}%
\providecommand \@@startlink[1]{}%
\providecommand \@@endlink[0]{}%
\providecommand \url  [0]{\begingroup\@sanitize@url \@url }%
\providecommand \@url [1]{\endgroup\@href {#1}{\urlprefix }}%
\providecommand \urlprefix  [0]{URL }%
\providecommand \Eprint [0]{\href }%
\providecommand \doibase [0]{https://doi.org/}%
\providecommand \selectlanguage [0]{\@gobble}%
\providecommand \bibinfo  [0]{\@secondoftwo}%
\providecommand \bibfield  [0]{\@secondoftwo}%
\providecommand \translation [1]{[#1]}%
\providecommand \BibitemOpen [0]{}%
\providecommand \bibitemStop [0]{}%
\providecommand \bibitemNoStop [0]{.\EOS\space}%
\providecommand \EOS [0]{\spacefactor3000\relax}%
\providecommand \BibitemShut  [1]{\csname bibitem#1\endcsname}%
\let\auto@bib@innerbib\@empty
%</preamble>
\bibitem [{\citenamefont {Albertazzi}\ \emph {et~al.}(2013)\citenamefont
  {Albertazzi}, \citenamefont {B\'{e}ard}, \citenamefont {Ciardi},
  \citenamefont {Vinci}, \citenamefont {Albrecht}, \citenamefont {Billette},
  \citenamefont {Burris-Mog}, \citenamefont {Chen}, \citenamefont {Da~Silva},
  \citenamefont {Dittrich}, \citenamefont {Herrmannsd\"{o}rfer}, \citenamefont
  {Hirardin}, \citenamefont {Kroll}, \citenamefont {Nakatsutsumi},
  \citenamefont {Nitsche}, \citenamefont {Riconda}, \citenamefont {Romagnagni},
  \citenamefont {Schlenvoigt}, \citenamefont {Simond}, \citenamefont
  {Veuillot}, \citenamefont {Cowan}, \citenamefont {Portugall}, \citenamefont
  {Pépin},\ and\ \citenamefont {Fuchs}}]{AlbertazziRSI2013}%
  \BibitemOpen
  \bibfield  {author} {\bibinfo {author} {\bibfnamefont {B.}~\bibnamefont
  {Albertazzi}}, \bibinfo {author} {\bibfnamefont {J.}~\bibnamefont
  {B\'{e}ard}}, \bibinfo {author} {\bibfnamefont {A.}~\bibnamefont {Ciardi}},
  \bibinfo {author} {\bibfnamefont {T.}~\bibnamefont {Vinci}}, \bibinfo
  {author} {\bibfnamefont {J.}~\bibnamefont {Albrecht}}, \bibinfo {author}
  {\bibfnamefont {J.}~\bibnamefont {Billette}}, \bibinfo {author}
  {\bibfnamefont {T.}~\bibnamefont {Burris-Mog}}, \bibinfo {author}
  {\bibfnamefont {S.~N.}\ \bibnamefont {Chen}}, \bibinfo {author}
  {\bibfnamefont {D.}~\bibnamefont {Da~Silva}}, \bibinfo {author}
  {\bibfnamefont {S.}~\bibnamefont {Dittrich}}, \bibinfo {author}
  {\bibfnamefont {T.}~\bibnamefont {Herrmannsd\"{o}rfer}}, \bibinfo {author}
  {\bibfnamefont {B.}~\bibnamefont {Hirardin}}, \bibinfo {author}
  {\bibfnamefont {F.}~\bibnamefont {Kroll}}, \bibinfo {author} {\bibfnamefont
  {M.}~\bibnamefont {Nakatsutsumi}}, \bibinfo {author} {\bibfnamefont
  {S.}~\bibnamefont {Nitsche}}, \bibinfo {author} {\bibfnamefont
  {C.}~\bibnamefont {Riconda}}, \bibinfo {author} {\bibfnamefont
  {L.}~\bibnamefont {Romagnagni}}, \bibinfo {author} {\bibfnamefont {H.-P.}\
  \bibnamefont {Schlenvoigt}}, \bibinfo {author} {\bibfnamefont
  {S.}~\bibnamefont {Simond}}, \bibinfo {author} {\bibfnamefont
  {E.}~\bibnamefont {Veuillot}}, \bibinfo {author} {\bibfnamefont {T.~E.}\
  \bibnamefont {Cowan}}, \bibinfo {author} {\bibfnamefont {O.}~\bibnamefont
  {Portugall}}, \bibinfo {author} {\bibfnamefont {H.}~\bibnamefont {Pépin}},\
  and\ \bibinfo {author} {\bibfnamefont {J.}~\bibnamefont {Fuchs}},\ }\bibfield
   {title} {\bibinfo {title} {Production of large volume, strongly magnetized
  laser-produced plasmas by use of pulsed external magnetic fields},\ }\href
  {https://doi.org/10.1063/1.4795551} {\bibfield  {journal} {\bibinfo
  {journal} {Rev. Sci. Instrum.}\ }\textbf {\bibinfo {volume} {84}},\ \bibinfo
  {pages} {043505} (\bibinfo {year} {2013})}\BibitemShut {NoStop}%
\bibitem [{\citenamefont {Albertazzi}\ \emph {et~al.}(2014)\citenamefont
  {Albertazzi}, \citenamefont {Ciardi}, \citenamefont {Nakatsutsumi},
  \citenamefont {Vinci}, \citenamefont {B\'{e}ard}, \citenamefont {Bonito},
  \citenamefont {Billette}, \citenamefont {Borghesi}, \citenamefont {Burkley},
  \citenamefont {Chen}, \citenamefont {Cowan}, \citenamefont
  {Herrmannsd\"{o}rfer}, \citenamefont {Higginson}, \citenamefont {Kroll},
  \citenamefont {Pikuz}, \citenamefont {Naughton}, \citenamefont {Romagnani},
  \citenamefont {Riconda}, \citenamefont {Revet}, \citenamefont {Riquier},
  \citenamefont {Schlenvoigt}, \citenamefont {Skobelev}, \citenamefont
  {Faenov}, \citenamefont {Soloviev}, \citenamefont {Huarte-Espinosa},
  \citenamefont {Frank}, \citenamefont {Portugall}, \citenamefont {Pépin},\
  and\ \citenamefont {Fuchs}}]{AlbertazziScience2014}%
  \BibitemOpen
  \bibfield  {author} {\bibinfo {author} {\bibfnamefont {B.}~\bibnamefont
  {Albertazzi}}, \bibinfo {author} {\bibfnamefont {A.}~\bibnamefont {Ciardi}},
  \bibinfo {author} {\bibfnamefont {M.}~\bibnamefont {Nakatsutsumi}}, \bibinfo
  {author} {\bibfnamefont {T.}~\bibnamefont {Vinci}}, \bibinfo {author}
  {\bibfnamefont {J.}~\bibnamefont {B\'{e}ard}}, \bibinfo {author}
  {\bibfnamefont {R.}~\bibnamefont {Bonito}}, \bibinfo {author} {\bibfnamefont
  {J.}~\bibnamefont {Billette}}, \bibinfo {author} {\bibfnamefont
  {M.}~\bibnamefont {Borghesi}}, \bibinfo {author} {\bibfnamefont
  {Z.}~\bibnamefont {Burkley}}, \bibinfo {author} {\bibfnamefont {S.~N.}\
  \bibnamefont {Chen}}, \bibinfo {author} {\bibfnamefont {T.~E.}\ \bibnamefont
  {Cowan}}, \bibinfo {author} {\bibfnamefont {T.}~\bibnamefont
  {Herrmannsd\"{o}rfer}}, \bibinfo {author} {\bibfnamefont {D.~P.}\
  \bibnamefont {Higginson}}, \bibinfo {author} {\bibfnamefont {F.}~\bibnamefont
  {Kroll}}, \bibinfo {author} {\bibfnamefont {S.~A.}\ \bibnamefont {Pikuz}},
  \bibinfo {author} {\bibfnamefont {K.}~\bibnamefont {Naughton}}, \bibinfo
  {author} {\bibfnamefont {L.}~\bibnamefont {Romagnani}}, \bibinfo {author}
  {\bibfnamefont {C.}~\bibnamefont {Riconda}}, \bibinfo {author} {\bibfnamefont
  {G.}~\bibnamefont {Revet}}, \bibinfo {author} {\bibfnamefont
  {R.}~\bibnamefont {Riquier}}, \bibinfo {author} {\bibfnamefont {H.-P.}\
  \bibnamefont {Schlenvoigt}}, \bibinfo {author} {\bibfnamefont {I.~Y.}\
  \bibnamefont {Skobelev}}, \bibinfo {author} {\bibfnamefont {A.}~\bibnamefont
  {Faenov}}, \bibinfo {author} {\bibfnamefont {A.}~\bibnamefont {Soloviev}},
  \bibinfo {author} {\bibfnamefont {M.}~\bibnamefont {Huarte-Espinosa}},
  \bibinfo {author} {\bibfnamefont {A.}~\bibnamefont {Frank}}, \bibinfo
  {author} {\bibfnamefont {O.}~\bibnamefont {Portugall}}, \bibinfo {author}
  {\bibfnamefont {H.}~\bibnamefont {Pépin}},\ and\ \bibinfo {author}
  {\bibfnamefont {J.}~\bibnamefont {Fuchs}},\ }\bibfield  {title} {\bibinfo
  {title} {Laboratory formation of a scaled protostellar jet by coaligned
  poloidal magnetic field},\ }\href {https://doi.org/10.1126/science.1259694}
  {\bibfield  {journal} {\bibinfo  {journal} {Science}\ }\textbf {\bibinfo
  {volume} {346}},\ \bibinfo {pages} {325} (\bibinfo {year}
  {2014})}\BibitemShut {NoStop}%
\bibitem [{\citenamefont {Mabey}\ \emph {et~al.}(2020)\citenamefont {Mabey},
  \citenamefont {Albertazzi}, \citenamefont {Rigon}, \citenamefont {Marquès},
  \citenamefont {Palmer}, \citenamefont {Topp-Mugglestone}, \citenamefont
  {Perez-Martin}, \citenamefont {Kroll}, \citenamefont {Brack}, \citenamefont
  {Cowan}, \citenamefont {Schramm}, \citenamefont {Falk}, \citenamefont
  {Gregori}, \citenamefont {Falize},\ and\ \citenamefont {Koenig}}]{Mabey2020}%
  \BibitemOpen
  \bibfield  {author} {\bibinfo {author} {\bibfnamefont {P.}~\bibnamefont
  {Mabey}}, \bibinfo {author} {\bibfnamefont {B.}~\bibnamefont {Albertazzi}},
  \bibinfo {author} {\bibfnamefont {G.}~\bibnamefont {Rigon}}, \bibinfo
  {author} {\bibfnamefont {J.-R.}\ \bibnamefont {Marquès}}, \bibinfo {author}
  {\bibfnamefont {C.~A.~J.}\ \bibnamefont {Palmer}}, \bibinfo {author}
  {\bibfnamefont {J.}~\bibnamefont {Topp-Mugglestone}}, \bibinfo {author}
  {\bibfnamefont {P.}~\bibnamefont {Perez-Martin}}, \bibinfo {author}
  {\bibfnamefont {F.}~\bibnamefont {Kroll}}, \bibinfo {author} {\bibfnamefont
  {F.~E.}\ \bibnamefont {Brack}}, \bibinfo {author} {\bibfnamefont {T.~E.}\
  \bibnamefont {Cowan}}, \bibinfo {author} {\bibfnamefont {U.}~\bibnamefont
  {Schramm}}, \bibinfo {author} {\bibfnamefont {K.}~\bibnamefont {Falk}},
  \bibinfo {author} {\bibfnamefont {G.}~\bibnamefont {Gregori}}, \bibinfo
  {author} {\bibfnamefont {E.}~\bibnamefont {Falize}},\ and\ \bibinfo {author}
  {\bibfnamefont {M.}~\bibnamefont {Koenig}},\ }\bibfield  {title} {\bibinfo
  {title} {Laboratory study of bilateral supernova remnants and continuous mhd
  shocks},\ }\href {https://doi.org/10.3847/1538-4357/ab92a4} {\bibfield
  {journal} {\bibinfo  {journal} {The Astrophysical Journal}\ }\textbf
  {\bibinfo {volume} {896}},\ \bibinfo {pages} {167} (\bibinfo {year}
  {2020})}\BibitemShut {NoStop}%
\bibitem [{\citenamefont {Moody}(2021)}]{Moody2021}%
  \BibitemOpen
  \bibfield  {author} {\bibinfo {author} {\bibfnamefont {J.~D.}\ \bibnamefont
  {Moody}},\ }\bibfield  {title} {\bibinfo {title} {Boosting
  inertial-confinement-fusion yield with magnetized fuel},\ }\href
  {https://doi.org/10.1103/Physics.14.51} {\bibfield  {journal} {\bibinfo
  {journal} {Physics}\ }\textbf {\bibinfo {volume} {14}},\ \bibinfo {pages}
  {51} (\bibinfo {year} {2021})}\BibitemShut {NoStop}%
\bibitem [{\citenamefont {Ishikawa}\ \emph {et~al.}(2012)\citenamefont
  {Ishikawa}, \citenamefont {Aoyagi}, \citenamefont {Asaka}, \citenamefont
  {Asano}, \citenamefont {Azumi}, \citenamefont {Bizen}, \citenamefont {Ego},
  \citenamefont {Fukami}, \citenamefont {Fukui}, \citenamefont {Furukawa},
  \citenamefont {Goto}, \citenamefont {Hanaki}, \citenamefont {Hara},
  \citenamefont {Hasegawa}, \citenamefont {Hatsui}, \citenamefont {Higashiya},
  \citenamefont {Hirono}, \citenamefont {Hosoda}, \citenamefont {Ishii},
  \citenamefont {Inagaki}, \citenamefont {Inubushi}, \citenamefont {Itoga},
  \citenamefont {Joti}, \citenamefont {Kago}, \citenamefont {Kameshima},
  \citenamefont {Kimura}, \citenamefont {Kirihara}, \citenamefont {Kiyomichi},
  \citenamefont {Kobayashi}, \citenamefont {Kondo}, \citenamefont {Kudo},
  \citenamefont {Maesaka}, \citenamefont {Mar{\'e}chal}, \citenamefont
  {Masuda}, \citenamefont {Matsubara}, \citenamefont {Matsumoto}, \citenamefont
  {Matsushita}, \citenamefont {Matsui}, \citenamefont {Nagasono}, \citenamefont
  {Nariyama}, \citenamefont {Ohashi}, \citenamefont {Ohata}, \citenamefont
  {Ohshima}, \citenamefont {Ono}, \citenamefont {Otake}, \citenamefont {Saji},
  \citenamefont {Sakurai}, \citenamefont {Sato}, \citenamefont {Sawada},
  \citenamefont {Seike}, \citenamefont {Shirasawa}, \citenamefont {Sugimoto},
  \citenamefont {Suzuki}, \citenamefont {Takahashi}, \citenamefont {Takebe},
  \citenamefont {Takeshita}, \citenamefont {Tamasaku}, \citenamefont {Tanaka},
  \citenamefont {Tanaka}, \citenamefont {Tanaka}, \citenamefont {Togashi},
  \citenamefont {Togawa}, \citenamefont {Tokuhisa}, \citenamefont {Tomizawa},
  \citenamefont {Tono}, \citenamefont {Wu}, \citenamefont {Yabashi},
  \citenamefont {Yamaga}, \citenamefont {Yamashita}, \citenamefont {Yanagida},
  \citenamefont {Zhang}, \citenamefont {Shintake}, \citenamefont {Kitamura},\
  and\ \citenamefont {Kumagai}}]{IshikawaNP2012}%
  \BibitemOpen
  \bibfield  {author} {\bibinfo {author} {\bibfnamefont {T.}~\bibnamefont
  {Ishikawa}}, \bibinfo {author} {\bibfnamefont {H.}~\bibnamefont {Aoyagi}},
  \bibinfo {author} {\bibfnamefont {T.}~\bibnamefont {Asaka}}, \bibinfo
  {author} {\bibfnamefont {Y.}~\bibnamefont {Asano}}, \bibinfo {author}
  {\bibfnamefont {N.}~\bibnamefont {Azumi}}, \bibinfo {author} {\bibfnamefont
  {T.}~\bibnamefont {Bizen}}, \bibinfo {author} {\bibfnamefont
  {H.}~\bibnamefont {Ego}}, \bibinfo {author} {\bibfnamefont {K.}~\bibnamefont
  {Fukami}}, \bibinfo {author} {\bibfnamefont {T.}~\bibnamefont {Fukui}},
  \bibinfo {author} {\bibfnamefont {Y.}~\bibnamefont {Furukawa}}, \bibinfo
  {author} {\bibfnamefont {S.}~\bibnamefont {Goto}}, \bibinfo {author}
  {\bibfnamefont {H.}~\bibnamefont {Hanaki}}, \bibinfo {author} {\bibfnamefont
  {T.}~\bibnamefont {Hara}}, \bibinfo {author} {\bibfnamefont {T.}~\bibnamefont
  {Hasegawa}}, \bibinfo {author} {\bibfnamefont {T.}~\bibnamefont {Hatsui}},
  \bibinfo {author} {\bibfnamefont {A.}~\bibnamefont {Higashiya}}, \bibinfo
  {author} {\bibfnamefont {T.}~\bibnamefont {Hirono}}, \bibinfo {author}
  {\bibfnamefont {N.}~\bibnamefont {Hosoda}}, \bibinfo {author} {\bibfnamefont
  {M.}~\bibnamefont {Ishii}}, \bibinfo {author} {\bibfnamefont
  {T.}~\bibnamefont {Inagaki}}, \bibinfo {author} {\bibfnamefont
  {Y.}~\bibnamefont {Inubushi}}, \bibinfo {author} {\bibfnamefont
  {T.}~\bibnamefont {Itoga}}, \bibinfo {author} {\bibfnamefont
  {Y.}~\bibnamefont {Joti}}, \bibinfo {author} {\bibfnamefont {M.}~\bibnamefont
  {Kago}}, \bibinfo {author} {\bibfnamefont {T.}~\bibnamefont {Kameshima}},
  \bibinfo {author} {\bibfnamefont {H.}~\bibnamefont {Kimura}}, \bibinfo
  {author} {\bibfnamefont {Y.}~\bibnamefont {Kirihara}}, \bibinfo {author}
  {\bibfnamefont {A.}~\bibnamefont {Kiyomichi}}, \bibinfo {author}
  {\bibfnamefont {T.}~\bibnamefont {Kobayashi}}, \bibinfo {author}
  {\bibfnamefont {C.}~\bibnamefont {Kondo}}, \bibinfo {author} {\bibfnamefont
  {T.}~\bibnamefont {Kudo}}, \bibinfo {author} {\bibfnamefont {H.}~\bibnamefont
  {Maesaka}}, \bibinfo {author} {\bibfnamefont {X.~M.}\ \bibnamefont
  {Mar{\'e}chal}}, \bibinfo {author} {\bibfnamefont {T.}~\bibnamefont
  {Masuda}}, \bibinfo {author} {\bibfnamefont {S.}~\bibnamefont {Matsubara}},
  \bibinfo {author} {\bibfnamefont {T.}~\bibnamefont {Matsumoto}}, \bibinfo
  {author} {\bibfnamefont {T.}~\bibnamefont {Matsushita}}, \bibinfo {author}
  {\bibfnamefont {S.}~\bibnamefont {Matsui}}, \bibinfo {author} {\bibfnamefont
  {M.}~\bibnamefont {Nagasono}}, \bibinfo {author} {\bibfnamefont
  {N.}~\bibnamefont {Nariyama}}, \bibinfo {author} {\bibfnamefont
  {H.}~\bibnamefont {Ohashi}}, \bibinfo {author} {\bibfnamefont
  {T.}~\bibnamefont {Ohata}}, \bibinfo {author} {\bibfnamefont
  {T.}~\bibnamefont {Ohshima}}, \bibinfo {author} {\bibfnamefont
  {S.}~\bibnamefont {Ono}}, \bibinfo {author} {\bibfnamefont {Y.}~\bibnamefont
  {Otake}}, \bibinfo {author} {\bibfnamefont {C.}~\bibnamefont {Saji}},
  \bibinfo {author} {\bibfnamefont {T.}~\bibnamefont {Sakurai}}, \bibinfo
  {author} {\bibfnamefont {T.}~\bibnamefont {Sato}}, \bibinfo {author}
  {\bibfnamefont {K.}~\bibnamefont {Sawada}}, \bibinfo {author} {\bibfnamefont
  {T.}~\bibnamefont {Seike}}, \bibinfo {author} {\bibfnamefont
  {K.}~\bibnamefont {Shirasawa}}, \bibinfo {author} {\bibfnamefont
  {T.}~\bibnamefont {Sugimoto}}, \bibinfo {author} {\bibfnamefont
  {S.}~\bibnamefont {Suzuki}}, \bibinfo {author} {\bibfnamefont
  {S.}~\bibnamefont {Takahashi}}, \bibinfo {author} {\bibfnamefont
  {H.}~\bibnamefont {Takebe}}, \bibinfo {author} {\bibfnamefont
  {K.}~\bibnamefont {Takeshita}}, \bibinfo {author} {\bibfnamefont
  {K.}~\bibnamefont {Tamasaku}}, \bibinfo {author} {\bibfnamefont
  {H.}~\bibnamefont {Tanaka}}, \bibinfo {author} {\bibfnamefont
  {R.}~\bibnamefont {Tanaka}}, \bibinfo {author} {\bibfnamefont
  {T.}~\bibnamefont {Tanaka}}, \bibinfo {author} {\bibfnamefont
  {T.}~\bibnamefont {Togashi}}, \bibinfo {author} {\bibfnamefont
  {K.}~\bibnamefont {Togawa}}, \bibinfo {author} {\bibfnamefont
  {A.}~\bibnamefont {Tokuhisa}}, \bibinfo {author} {\bibfnamefont
  {H.}~\bibnamefont {Tomizawa}}, \bibinfo {author} {\bibfnamefont
  {K.}~\bibnamefont {Tono}}, \bibinfo {author} {\bibfnamefont {S.}~\bibnamefont
  {Wu}}, \bibinfo {author} {\bibfnamefont {M.}~\bibnamefont {Yabashi}},
  \bibinfo {author} {\bibfnamefont {M.}~\bibnamefont {Yamaga}}, \bibinfo
  {author} {\bibfnamefont {A.}~\bibnamefont {Yamashita}}, \bibinfo {author}
  {\bibfnamefont {K.}~\bibnamefont {Yanagida}}, \bibinfo {author}
  {\bibfnamefont {C.}~\bibnamefont {Zhang}}, \bibinfo {author} {\bibfnamefont
  {T.}~\bibnamefont {Shintake}}, \bibinfo {author} {\bibfnamefont
  {H.}~\bibnamefont {Kitamura}},\ and\ \bibinfo {author} {\bibfnamefont
  {N.}~\bibnamefont {Kumagai}},\ }\bibfield  {title} {\bibinfo {title} {\rm{A
  compact X-ray free-electron laser emitting in the sub-{\aa}ngstr{\"o}m
  region}},\ }\href {https://doi.org/10.1038/nphoton.2012.141} {\bibfield
  {journal} {\bibinfo  {journal} {Nat. Photonics}\ }\textbf {\bibinfo {volume}
  {6}},\ \bibinfo {pages} {540} (\bibinfo {year} {2012})}\BibitemShut {NoStop}%
\bibitem [{\citenamefont {Inubushi}\ \emph {et~al.}(2020)\citenamefont
  {Inubushi}, \citenamefont {Yabuuchi}, \citenamefont {Togashi}, \citenamefont
  {Sueda}, \citenamefont {Miyanishi}, \citenamefont {Tange}, \citenamefont
  {Ozaki}, \citenamefont {Matsuoka}, \citenamefont {Kodama}, \citenamefont
  {Osaka}, \citenamefont {Matsuyama}, \citenamefont {Yamauchi}, \citenamefont
  {Yumoto}, \citenamefont {Koyama}, \citenamefont {Ohashi}, \citenamefont
  {Tono},\ and\ \citenamefont {Yabashi}}]{Inubushi2020AS}%
  \BibitemOpen
  \bibfield  {author} {\bibinfo {author} {\bibfnamefont {Y.}~\bibnamefont
  {Inubushi}}, \bibinfo {author} {\bibfnamefont {T.}~\bibnamefont {Yabuuchi}},
  \bibinfo {author} {\bibfnamefont {T.}~\bibnamefont {Togashi}}, \bibinfo
  {author} {\bibfnamefont {K.}~\bibnamefont {Sueda}}, \bibinfo {author}
  {\bibfnamefont {K.}~\bibnamefont {Miyanishi}}, \bibinfo {author}
  {\bibfnamefont {Y.}~\bibnamefont {Tange}}, \bibinfo {author} {\bibfnamefont
  {N.}~\bibnamefont {Ozaki}}, \bibinfo {author} {\bibfnamefont
  {T.}~\bibnamefont {Matsuoka}}, \bibinfo {author} {\bibfnamefont
  {R.}~\bibnamefont {Kodama}}, \bibinfo {author} {\bibfnamefont
  {T.}~\bibnamefont {Osaka}}, \bibinfo {author} {\bibfnamefont
  {S.}~\bibnamefont {Matsuyama}}, \bibinfo {author} {\bibfnamefont
  {K.}~\bibnamefont {Yamauchi}}, \bibinfo {author} {\bibfnamefont
  {H.}~\bibnamefont {Yumoto}}, \bibinfo {author} {\bibfnamefont
  {T.}~\bibnamefont {Koyama}}, \bibinfo {author} {\bibfnamefont
  {H.}~\bibnamefont {Ohashi}}, \bibinfo {author} {\bibfnamefont
  {K.}~\bibnamefont {Tono}},\ and\ \bibinfo {author} {\bibfnamefont
  {M.}~\bibnamefont {Yabashi}},\ }\bibfield  {title} {\bibinfo {title}
  {Development of an experimental platform for combinative use of an xfel and a
  high-power nanosecond laser},\ }\href {https://doi.org/10.3390/app10072224}
  {\bibfield  {journal} {\bibinfo  {journal} {Appl. Sci.}\ }\textbf {\bibinfo
  {volume} {10}},\ \bibinfo {pages} {2224} (\bibinfo {year}
  {2020})}\BibitemShut {NoStop}%
\bibitem [{\citenamefont {Faenov}\ \emph {et~al.}(2018)\citenamefont {Faenov},
  \citenamefont {Pikuz}, \citenamefont {Mabey}, \citenamefont {Albertazzi},
  \citenamefont {Michel}, \citenamefont {Rigon}, \citenamefont {Pikuz},
  \citenamefont {Buzmakov}, \citenamefont {Makarov}, \citenamefont {Ozaki},
  \citenamefont {Matsuoka}, \citenamefont {Katagiri}, \citenamefont
  {Miyanishi}, \citenamefont {Takahashi}, \citenamefont {Tanaka}, \citenamefont
  {Inubushi}, \citenamefont {Togashi}, \citenamefont {Yabuuchi}, \citenamefont
  {Yabashi}, \citenamefont {Casner}, \citenamefont {Kodama},\ and\
  \citenamefont {Koenig}}]{Faenov2018}%
  \BibitemOpen
  \bibfield  {author} {\bibinfo {author} {\bibfnamefont {A.~Y.}\ \bibnamefont
  {Faenov}}, \bibinfo {author} {\bibfnamefont {T.~A.}\ \bibnamefont {Pikuz}},
  \bibinfo {author} {\bibfnamefont {P.}~\bibnamefont {Mabey}}, \bibinfo
  {author} {\bibfnamefont {B.}~\bibnamefont {Albertazzi}}, \bibinfo {author}
  {\bibfnamefont {T.}~\bibnamefont {Michel}}, \bibinfo {author} {\bibfnamefont
  {G.}~\bibnamefont {Rigon}}, \bibinfo {author} {\bibfnamefont {S.~A.}\
  \bibnamefont {Pikuz}}, \bibinfo {author} {\bibfnamefont {A.}~\bibnamefont
  {Buzmakov}}, \bibinfo {author} {\bibfnamefont {S.}~\bibnamefont {Makarov}},
  \bibinfo {author} {\bibfnamefont {N.}~\bibnamefont {Ozaki}}, \bibinfo
  {author} {\bibfnamefont {T.}~\bibnamefont {Matsuoka}}, \bibinfo {author}
  {\bibfnamefont {K.}~\bibnamefont {Katagiri}}, \bibinfo {author}
  {\bibfnamefont {K.}~\bibnamefont {Miyanishi}}, \bibinfo {author}
  {\bibfnamefont {K.}~\bibnamefont {Takahashi}}, \bibinfo {author}
  {\bibfnamefont {K.~A.}\ \bibnamefont {Tanaka}}, \bibinfo {author}
  {\bibfnamefont {Y.}~\bibnamefont {Inubushi}}, \bibinfo {author}
  {\bibfnamefont {T.}~\bibnamefont {Togashi}}, \bibinfo {author} {\bibfnamefont
  {T.}~\bibnamefont {Yabuuchi}}, \bibinfo {author} {\bibfnamefont
  {M.}~\bibnamefont {Yabashi}}, \bibinfo {author} {\bibfnamefont
  {A.}~\bibnamefont {Casner}}, \bibinfo {author} {\bibfnamefont
  {R.}~\bibnamefont {Kodama}},\ and\ \bibinfo {author} {\bibfnamefont
  {M.}~\bibnamefont {Koenig}},\ }\bibfield  {title} {\bibinfo {title} {Advanced
  high resolution x-ray diagnostic for hedp experiments},\ }\href
  {https://doi.org/10.1038/s41598-018-34717-9} {\bibfield  {journal} {\bibinfo
  {journal} {Sci. Rep.}\ }\textbf {\bibinfo {volume} {8}},\ \bibinfo {pages}
  {16407} (\bibinfo {year} {2018})}\BibitemShut {NoStop}%
\bibitem [{\citenamefont {Rigon}\ \emph {et~al.}(2021)\citenamefont {Rigon},
  \citenamefont {Albertazzi}, \citenamefont {Pikuz}, \citenamefont {Mabey},
  \citenamefont {Michel}, \citenamefont {Marquès}, \citenamefont {Palmer},
  \citenamefont {Topp-Mugglestone}, \citenamefont {Perez-Martin}, \citenamefont
  {Kroll}, \citenamefont {Brack}, \citenamefont {Cowan}, \citenamefont
  {Schramm}, \citenamefont {Falk}, \citenamefont {Gregori}, \citenamefont
  {Falize},\ and\ \citenamefont {Koenig}}]{Rigon2021}%
  \BibitemOpen
  \bibfield  {author} {\bibinfo {author} {\bibfnamefont {G.}~\bibnamefont
  {Rigon}}, \bibinfo {author} {\bibfnamefont {B.}~\bibnamefont {Albertazzi}},
  \bibinfo {author} {\bibfnamefont {T.}~\bibnamefont {Pikuz}}, \bibinfo
  {author} {\bibfnamefont {P.}~\bibnamefont {Mabey}}, \bibinfo {author}
  {\bibfnamefont {T.}~\bibnamefont {Michel}}, \bibinfo {author} {\bibfnamefont
  {J.-R.}\ \bibnamefont {Marquès}}, \bibinfo {author} {\bibfnamefont
  {C.~A.~J.}\ \bibnamefont {Palmer}}, \bibinfo {author} {\bibfnamefont
  {J.}~\bibnamefont {Topp-Mugglestone}}, \bibinfo {author} {\bibfnamefont
  {P.}~\bibnamefont {Perez-Martin}}, \bibinfo {author} {\bibfnamefont
  {F.}~\bibnamefont {Kroll}}, \bibinfo {author} {\bibfnamefont {F.~E.}\
  \bibnamefont {Brack}}, \bibinfo {author} {\bibfnamefont {T.~E.}\ \bibnamefont
  {Cowan}}, \bibinfo {author} {\bibfnamefont {U.}~\bibnamefont {Schramm}},
  \bibinfo {author} {\bibfnamefont {K.}~\bibnamefont {Falk}}, \bibinfo {author}
  {\bibfnamefont {G.}~\bibnamefont {Gregori}}, \bibinfo {author} {\bibfnamefont
  {E.}~\bibnamefont {Falize}},\ and\ \bibinfo {author} {\bibfnamefont
  {M.}~\bibnamefont {Koenig}},\ }\bibfield  {title} {\bibinfo {title}
  {Micron-scale phenomena observed in a turbulent laser-produced plasma},\
  }\href {https://doi.org/10.1038/s41467-021-22891-w} {\bibfield  {journal}
  {\bibinfo  {journal} {Nat. Commun.}\ }\textbf {\bibinfo {volume} {12}},\
  \bibinfo {pages} {2679} (\bibinfo {year} {2021})}\BibitemShut {NoStop}%
\bibitem [{\citenamefont {Katagiri}\ \emph {et~al.}(2023)\citenamefont
  {Katagiri}, \citenamefont {Pikuz}, \citenamefont {Fang}, \citenamefont
  {Albertazzi}, \citenamefont {Egashira}, \citenamefont {Inubushi},
  \citenamefont {Kamimura}, \citenamefont {Kodama}, \citenamefont {Koenig},
  \citenamefont {Kozioziemski}, \citenamefont {Masaoka}, \citenamefont
  {Miyanishi}, \citenamefont {Nakamura}, \citenamefont {Ota}, \citenamefont
  {Rigon}, \citenamefont {Sakawa}, \citenamefont {Sano}, \citenamefont
  {Schoofs}, \citenamefont {Smith}, \citenamefont {Sueda}, \citenamefont
  {Togashi}, \citenamefont {Vinci}, \citenamefont {Wang}, \citenamefont
  {Yabashi}, \citenamefont {Yabuuchi}, \citenamefont {Dresselhaus-Marais},\
  and\ \citenamefont {Ozaki}}]{Katagiri2023Science}%
  \BibitemOpen
  \bibfield  {author} {\bibinfo {author} {\bibfnamefont {K.}~\bibnamefont
  {Katagiri}}, \bibinfo {author} {\bibfnamefont {T.}~\bibnamefont {Pikuz}},
  \bibinfo {author} {\bibfnamefont {L.}~\bibnamefont {Fang}}, \bibinfo {author}
  {\bibfnamefont {B.}~\bibnamefont {Albertazzi}}, \bibinfo {author}
  {\bibfnamefont {S.}~\bibnamefont {Egashira}}, \bibinfo {author}
  {\bibfnamefont {Y.}~\bibnamefont {Inubushi}}, \bibinfo {author}
  {\bibfnamefont {G.}~\bibnamefont {Kamimura}}, \bibinfo {author}
  {\bibfnamefont {R.}~\bibnamefont {Kodama}}, \bibinfo {author} {\bibfnamefont
  {M.}~\bibnamefont {Koenig}}, \bibinfo {author} {\bibfnamefont
  {B.}~\bibnamefont {Kozioziemski}}, \bibinfo {author} {\bibfnamefont
  {G.}~\bibnamefont {Masaoka}}, \bibinfo {author} {\bibfnamefont
  {K.}~\bibnamefont {Miyanishi}}, \bibinfo {author} {\bibfnamefont
  {H.}~\bibnamefont {Nakamura}}, \bibinfo {author} {\bibfnamefont
  {M.}~\bibnamefont {Ota}}, \bibinfo {author} {\bibfnamefont {G.}~\bibnamefont
  {Rigon}}, \bibinfo {author} {\bibfnamefont {Y.}~\bibnamefont {Sakawa}},
  \bibinfo {author} {\bibfnamefont {T.}~\bibnamefont {Sano}}, \bibinfo {author}
  {\bibfnamefont {F.}~\bibnamefont {Schoofs}}, \bibinfo {author} {\bibfnamefont
  {Z.~J.}\ \bibnamefont {Smith}}, \bibinfo {author} {\bibfnamefont
  {K.}~\bibnamefont {Sueda}}, \bibinfo {author} {\bibfnamefont
  {T.}~\bibnamefont {Togashi}}, \bibinfo {author} {\bibfnamefont
  {T.}~\bibnamefont {Vinci}}, \bibinfo {author} {\bibfnamefont
  {Y.}~\bibnamefont {Wang}}, \bibinfo {author} {\bibfnamefont {M.}~\bibnamefont
  {Yabashi}}, \bibinfo {author} {\bibfnamefont {T.}~\bibnamefont {Yabuuchi}},
  \bibinfo {author} {\bibfnamefont {L.~E.}\ \bibnamefont
  {Dresselhaus-Marais}},\ and\ \bibinfo {author} {\bibfnamefont
  {N.}~\bibnamefont {Ozaki}},\ }\bibfield  {title} {\bibinfo {title} {Transonic
  dislocation propagation in diamond},\ }\href
  {https://doi.org/10.1126/science.adg7840} {\bibfield  {journal} {\bibinfo
  {journal} {Science}\ }\textbf {\bibinfo {volume} {382}},\ \bibinfo {pages}
  {69} (\bibinfo {year} {2023})}\BibitemShut {NoStop}%
\bibitem [{\citenamefont {Albertazzi}\ \emph {et~al.}(2017)\citenamefont
  {Albertazzi}, \citenamefont {Ozaki}, \citenamefont {Zhakhovsky},
  \citenamefont {Faenov}, \citenamefont {Habara}, \citenamefont {Harmand},
  \citenamefont {Hartley}, \citenamefont {Ilnitsky}, \citenamefont {Inogamov},
  \citenamefont {Inubushi}, \citenamefont {Ishikawa}, \citenamefont {Katayama},
  \citenamefont {Koyama}, \citenamefont {Koenig}, \citenamefont {Krygier},
  \citenamefont {Matsuoka}, \citenamefont {Matsuyama}, \citenamefont {McBride},
  \citenamefont {Migdal}, \citenamefont {Morard}, \citenamefont {Ohashi},
  \citenamefont {Okuchi}, \citenamefont {Pikuz}, \citenamefont {Purevjav},
  \citenamefont {Sakata}, \citenamefont {Sano}, \citenamefont {Sato},
  \citenamefont {Sekine}, \citenamefont {Seto}, \citenamefont {Takahashi},
  \citenamefont {Tanaka}, \citenamefont {Tange}, \citenamefont {Togashi},
  \citenamefont {Tono}, \citenamefont {Umeda}, \citenamefont {Vinci},
  \citenamefont {Yabashi}, \citenamefont {Yabuuchi}, \citenamefont {Yamauchi},
  \citenamefont {Yumoto},\ and\ \citenamefont {Kodama}}]{AlbertazziSA2017}%
  \BibitemOpen
  \bibfield  {author} {\bibinfo {author} {\bibfnamefont {B.}~\bibnamefont
  {Albertazzi}}, \bibinfo {author} {\bibfnamefont {N.}~\bibnamefont {Ozaki}},
  \bibinfo {author} {\bibfnamefont {V.}~\bibnamefont {Zhakhovsky}}, \bibinfo
  {author} {\bibfnamefont {A.}~\bibnamefont {Faenov}}, \bibinfo {author}
  {\bibfnamefont {H.}~\bibnamefont {Habara}}, \bibinfo {author} {\bibfnamefont
  {M.}~\bibnamefont {Harmand}}, \bibinfo {author} {\bibfnamefont
  {N.}~\bibnamefont {Hartley}}, \bibinfo {author} {\bibfnamefont
  {D.}~\bibnamefont {Ilnitsky}}, \bibinfo {author} {\bibfnamefont
  {N.}~\bibnamefont {Inogamov}}, \bibinfo {author} {\bibfnamefont
  {Y.}~\bibnamefont {Inubushi}}, \bibinfo {author} {\bibfnamefont
  {T.}~\bibnamefont {Ishikawa}}, \bibinfo {author} {\bibfnamefont
  {T.}~\bibnamefont {Katayama}}, \bibinfo {author} {\bibfnamefont
  {T.}~\bibnamefont {Koyama}}, \bibinfo {author} {\bibfnamefont
  {M.}~\bibnamefont {Koenig}}, \bibinfo {author} {\bibfnamefont
  {A.}~\bibnamefont {Krygier}}, \bibinfo {author} {\bibfnamefont
  {T.}~\bibnamefont {Matsuoka}}, \bibinfo {author} {\bibfnamefont
  {S.}~\bibnamefont {Matsuyama}}, \bibinfo {author} {\bibfnamefont
  {E.}~\bibnamefont {McBride}}, \bibinfo {author} {\bibfnamefont {K.~P.}\
  \bibnamefont {Migdal}}, \bibinfo {author} {\bibfnamefont {G.}~\bibnamefont
  {Morard}}, \bibinfo {author} {\bibfnamefont {H.}~\bibnamefont {Ohashi}},
  \bibinfo {author} {\bibfnamefont {T.}~\bibnamefont {Okuchi}}, \bibinfo
  {author} {\bibfnamefont {T.}~\bibnamefont {Pikuz}}, \bibinfo {author}
  {\bibfnamefont {N.}~\bibnamefont {Purevjav}}, \bibinfo {author}
  {\bibfnamefont {O.}~\bibnamefont {Sakata}}, \bibinfo {author} {\bibfnamefont
  {Y.}~\bibnamefont {Sano}}, \bibinfo {author} {\bibfnamefont {T.}~\bibnamefont
  {Sato}}, \bibinfo {author} {\bibfnamefont {T.}~\bibnamefont {Sekine}},
  \bibinfo {author} {\bibfnamefont {Y.}~\bibnamefont {Seto}}, \bibinfo {author}
  {\bibfnamefont {K.}~\bibnamefont {Takahashi}}, \bibinfo {author}
  {\bibfnamefont {K.}~\bibnamefont {Tanaka}}, \bibinfo {author} {\bibfnamefont
  {Y.}~\bibnamefont {Tange}}, \bibinfo {author} {\bibfnamefont
  {T.}~\bibnamefont {Togashi}}, \bibinfo {author} {\bibfnamefont
  {K.}~\bibnamefont {Tono}}, \bibinfo {author} {\bibfnamefont {Y.}~\bibnamefont
  {Umeda}}, \bibinfo {author} {\bibfnamefont {T.}~\bibnamefont {Vinci}},
  \bibinfo {author} {\bibfnamefont {M.}~\bibnamefont {Yabashi}}, \bibinfo
  {author} {\bibfnamefont {T.}~\bibnamefont {Yabuuchi}}, \bibinfo {author}
  {\bibfnamefont {K.}~\bibnamefont {Yamauchi}}, \bibinfo {author}
  {\bibfnamefont {H.}~\bibnamefont {Yumoto}},\ and\ \bibinfo {author}
  {\bibfnamefont {R.}~\bibnamefont {Kodama}},\ }\bibfield  {title} {\bibinfo
  {title} {Dynamic fracture of tantalum under extreme tensile stress},\ }\href
  {https://doi.org/10.1126/sciadv.1602705} {\bibfield  {journal} {\bibinfo
  {journal} {Sci. Adv.}\ }\textbf {\bibinfo {volume} {3}},\ \bibinfo {pages}
  {e1602705} (\bibinfo {year} {2017})}\BibitemShut {NoStop}%
\bibitem [{\citenamefont {Katagiri}\ \emph {et~al.}(2020)\citenamefont
  {Katagiri}, \citenamefont {Ozaki}, \citenamefont {Umeda}, \citenamefont
  {Irifune}, \citenamefont {Kamimura}, \citenamefont {Miyanishi}, \citenamefont
  {Sano}, \citenamefont {Sekine},\ and\ \citenamefont
  {Kodama}}]{Katagiri2020PRL}%
  \BibitemOpen
  \bibfield  {author} {\bibinfo {author} {\bibfnamefont {K.}~\bibnamefont
  {Katagiri}}, \bibinfo {author} {\bibfnamefont {N.}~\bibnamefont {Ozaki}},
  \bibinfo {author} {\bibfnamefont {Y.}~\bibnamefont {Umeda}}, \bibinfo
  {author} {\bibfnamefont {T.}~\bibnamefont {Irifune}}, \bibinfo {author}
  {\bibfnamefont {N.}~\bibnamefont {Kamimura}}, \bibinfo {author}
  {\bibfnamefont {K.}~\bibnamefont {Miyanishi}}, \bibinfo {author}
  {\bibfnamefont {T.}~\bibnamefont {Sano}}, \bibinfo {author} {\bibfnamefont
  {T.}~\bibnamefont {Sekine}},\ and\ \bibinfo {author} {\bibfnamefont
  {R.}~\bibnamefont {Kodama}},\ }\bibfield  {title} {\bibinfo {title} {Shock
  response of full density nanopolycrystalline diamond},\ }\href
  {https://doi.org/10.1103/PhysRevLett.125.185701} {\bibfield  {journal}
  {\bibinfo  {journal} {Phys. Rev. Lett.}\ }\textbf {\bibinfo {volume} {125}},\
  \bibinfo {pages} {185701} (\bibinfo {year} {2020})}\BibitemShut {NoStop}%
\bibitem [{\citenamefont {Kameshima}\ \emph {et~al.}(2019)\citenamefont
  {Kameshima}, \citenamefont {Takeuchi}, \citenamefont {Uesugi}, \citenamefont
  {Kudo}, \citenamefont {Kohmura}, \citenamefont {Tamasaku}, \citenamefont
  {Muramatsu}, \citenamefont {Yanagitani}, \citenamefont {Yabashi},\ and\
  \citenamefont {Hatsui}}]{KameshimaOL2019}%
  \BibitemOpen
  \bibfield  {author} {\bibinfo {author} {\bibfnamefont {T.}~\bibnamefont
  {Kameshima}}, \bibinfo {author} {\bibfnamefont {A.}~\bibnamefont {Takeuchi}},
  \bibinfo {author} {\bibfnamefont {K.}~\bibnamefont {Uesugi}}, \bibinfo
  {author} {\bibfnamefont {T.}~\bibnamefont {Kudo}}, \bibinfo {author}
  {\bibfnamefont {Y.}~\bibnamefont {Kohmura}}, \bibinfo {author} {\bibfnamefont
  {K.}~\bibnamefont {Tamasaku}}, \bibinfo {author} {\bibfnamefont
  {K.}~\bibnamefont {Muramatsu}}, \bibinfo {author} {\bibfnamefont
  {T.}~\bibnamefont {Yanagitani}}, \bibinfo {author} {\bibfnamefont
  {M.}~\bibnamefont {Yabashi}},\ and\ \bibinfo {author} {\bibfnamefont
  {T.}~\bibnamefont {Hatsui}},\ }\bibfield  {title} {\bibinfo {title}
  {\rm{Development of an X-ray imaging detector to resolve 200 nm
  line-and-space patterns by using transparent ceramics layers bonded by
  solid-state diffusion}},\ }\href {https://doi.org/10.1364/OL.44.001403}
  {\bibfield  {journal} {\bibinfo  {journal} {Opt. Lett.}\ }\textbf {\bibinfo
  {volume} {44}},\ \bibinfo {pages} {1403} (\bibinfo {year}
  {2019})}\BibitemShut {NoStop}%
\bibitem [{\citenamefont {Ikeda}\ \emph {et~al.}(2024)\citenamefont {Ikeda},
  \citenamefont {Noda}, \citenamefont {Shimbori}, \citenamefont {Seki},
  \citenamefont {Bhoi}, \citenamefont {Ishita}, \citenamefont {Nakamura},
  \citenamefont {Matsubayashi},\ and\ \citenamefont {Akiba}}]{IkedaJAP2024}%
  \BibitemOpen
  \bibfield  {author} {\bibinfo {author} {\bibfnamefont {A.}~\bibnamefont
  {Ikeda}}, \bibinfo {author} {\bibfnamefont {K.}~\bibnamefont {Noda}},
  \bibinfo {author} {\bibfnamefont {K.}~\bibnamefont {Shimbori}}, \bibinfo
  {author} {\bibfnamefont {K.}~\bibnamefont {Seki}}, \bibinfo {author}
  {\bibfnamefont {D.}~\bibnamefont {Bhoi}}, \bibinfo {author} {\bibfnamefont
  {A.}~\bibnamefont {Ishita}}, \bibinfo {author} {\bibfnamefont
  {J.}~\bibnamefont {Nakamura}}, \bibinfo {author} {\bibfnamefont
  {K.}~\bibnamefont {Matsubayashi}},\ and\ \bibinfo {author} {\bibfnamefont
  {K.}~\bibnamefont {Akiba}},\ }\bibfield  {title} {\bibinfo {title} {\rm{A
  concise 40 T pulse magnet for condensed matter experiments}},\ }\href
  {https://doi.org/10.1063/5.0231640} {\bibfield  {journal} {\bibinfo
  {journal} {J. Appl. Phys.}\ }\textbf {\bibinfo {volume} {136}},\ \bibinfo
  {pages} {175902} (\bibinfo {year} {2024})}\BibitemShut {NoStop}%
\bibitem [{\citenamefont {Goldstein}(2001)}]{Goldstein2001}%
  \BibitemOpen
  \bibfield  {author} {\bibinfo {author} {\bibfnamefont {M.}~\bibnamefont
  {Goldstein}},\ }\bibfield  {title} {\bibinfo {title} {\rm{Major Unsolved
  Problems in Space Plasma Physics}},\ }\href
  {https://doi.org/10.1023/A:1012264131485} {\bibfield  {journal} {\bibinfo
  {journal} {Astrophys. Space Sci.}\ }\textbf {\bibinfo {volume} {277}},\
  \bibinfo {pages} {349} (\bibinfo {year} {2001})}\BibitemShut {NoStop}%
\bibitem [{\citenamefont {Tu}\ and\ \citenamefont {Marsch}(1995)}]{Tu1995}%
  \BibitemOpen
  \bibfield  {author} {\bibinfo {author} {\bibfnamefont {C.~Y.}\ \bibnamefont
  {Tu}}\ and\ \bibinfo {author} {\bibfnamefont {E.}~\bibnamefont {Marsch}},\
  }\bibfield  {title} {\bibinfo {title} {\rm{MHD structures, waves and
  turbulence in the solar wind: Observations and theories}},\ }\href
  {https://doi.org/10.1007/BF00748891} {\bibfield  {journal} {\bibinfo
  {journal} {Space Sci. Rev.}\ }\textbf {\bibinfo {volume} {73}},\ \bibinfo
  {pages} {1} (\bibinfo {year} {1995})}\BibitemShut {NoStop}%
\bibitem [{\citenamefont {Sridhar}\ and\ \citenamefont
  {Goldreich}(1994)}]{Sridhar1994}%
  \BibitemOpen
  \bibfield  {author} {\bibinfo {author} {\bibfnamefont {S.}~\bibnamefont
  {Sridhar}}\ and\ \bibinfo {author} {\bibfnamefont {P.}~\bibnamefont
  {Goldreich}},\ }\bibfield  {title} {\bibinfo {title} {\rm{Toward a theory of
  interstellar turbulence. I: Weak Alfv\'{e}nic turbulence}},\ }\href
  {https://doi.org/10.1086/174600} {\bibfield  {journal} {\bibinfo  {journal}
  {Astrophys. J.}\ }\textbf {\bibinfo {volume} {432}},\ \bibinfo {pages} {612}
  (\bibinfo {year} {1994})}\BibitemShut {NoStop}%
\bibitem [{\citenamefont {Goldreich}\ and\ \citenamefont
  {Sridhar}(1995)}]{Goldreich1995}%
  \BibitemOpen
  \bibfield  {author} {\bibinfo {author} {\bibfnamefont {P.}~\bibnamefont
  {Goldreich}}\ and\ \bibinfo {author} {\bibfnamefont {S.}~\bibnamefont
  {Sridhar}},\ }\bibfield  {title} {\bibinfo {title} {\rm{Toward a theory of
  interstellar turbulence. 2: Strong Alfv\'{e}nic turbulence}},\ }\href
  {https://doi.org/10.1086/175121} {\bibfield  {journal} {\bibinfo  {journal}
  {Astrophys. J.}\ }\textbf {\bibinfo {volume} {438}},\ \bibinfo {pages} {763}
  (\bibinfo {year} {1995})}\BibitemShut {NoStop}%
\bibitem [{\citenamefont {Vishniac}\ and\ \citenamefont
  {Cho}(2000)}]{Vishniac2000}%
  \BibitemOpen
  \bibfield  {author} {\bibinfo {author} {\bibfnamefont {E.~T.}\ \bibnamefont
  {Vishniac}}\ and\ \bibinfo {author} {\bibfnamefont {J.}~\bibnamefont {Cho}},\
  }\bibfield  {title} {\bibinfo {title} {\rm{The Anisotropy of
  Magnetohydrodynamic Alfv\'{e}nic Turbulence}},\ }\href
  {https://doi.org/10.1086/309213} {\bibfield  {journal} {\bibinfo  {journal}
  {Astrophys. J.}\ }\textbf {\bibinfo {volume} {539}},\ \bibinfo {pages} {273}
  (\bibinfo {year} {2000})}\BibitemShut {NoStop}%
\bibitem [{\citenamefont {Schekochihin}(2022)}]{Schekochihin2022}%
  \BibitemOpen
  \bibfield  {author} {\bibinfo {author} {\bibfnamefont {A.~A.}\ \bibnamefont
  {Schekochihin}},\ }\bibfield  {title} {\bibinfo {title} {\rm{MHD Turbulence:
  A Biased Review}},\ }\href {https://doi.org/10.1017/S0022377822000721}
  {\bibfield  {journal} {\bibinfo  {journal} {J. Plasma Phys.}\ }\textbf
  {\bibinfo {volume} {88}},\ \bibinfo {pages} {Issue 5} (\bibinfo {year}
  {2022})}\BibitemShut {NoStop}%
\end{thebibliography}%

\end{document}